\documentclass[preprint2]{aastex}

\usepackage[colorlinks=true,citecolor=blue,breaklinks=true,linktocpage=true]{hyperref} %
\bibpunct{(}{)}{;}{a}{}{,} 
\usepackage[switch,pagewise]{lineno}
\usepackage{xspace}

\newcommand{\secref}[1]{Sect.~\ref{#1}}
\newcommand{\figref}[1]{Fig.~\ref{#1}}
\newcommand{\dateobs}{31 Aug 2013\xspace}
\newcommand{\arobs}{AR 11836\xspace}

\shorttitle{sunspot waves}
\shortauthors{Yuan et al.}

\begin{document}

\title{Oscillations in a sunspot with light bridges}
\author{Ding Yuan\altaffilmark{1,2,3}}
\email{Ding.Yuan@wis.kuleuven.be}
\author{Valery M. Nakariakov\altaffilmark{4,5,6}}
\author{Zhenghua Huang\altaffilmark{2}}
\author{Bo Li\altaffilmark{2}}
\author{Jiangtao Su\altaffilmark{1}}
\author{Yihua Yan\altaffilmark{1}}
\and
\author{Baolin Tan\altaffilmark{1}}
\altaffiltext{1}{Key Laboratory of Solar Activity, National Astronomical Observatories, Chinese Academy of Sciences, Beijing, 100012}
\altaffiltext{2}{School of Space Science and Physics, Shandong University, Weihai 246209, China}
\altaffiltext{3}{Centre for mathematical Plasma Astrophysics, Department of Mathematics, KU Leuven, Celestijnenlaan 200B bus 2400, B-3001 Leuven, Belgium}
\altaffiltext{4}{Centre for Fusion, Space and Astrophysics, Department of Physics, University of Warwick, Coventry CV4 7AL, UK}
\altaffiltext{5}{School of Space Research, Kyung Hee University, Yongin, 446-701, Gyeonggi, Korea}
\altaffiltext{6}{Central Astronomical Observatory of the Russian Academy of Sciences at Pulkovo, 196140 St Petersburg, Russia}

\begin{abstract}
Solar Optical Telescope onboard Hinode observed a sunspot (\arobs) with two light bridges (LBs) on \dateobs. We analysed a 2-hour \ion{Ca}{2} H emission intensity data set and detected strong 5-min oscillation power on both LBs and in the inner penumbra. The time-distance plot reveals that 5-min oscillation phase does not vary significantly along the thin bridge, indicating that the oscillations are likely to originate from the underneath. The slit taken along the central axis of the wide light bridge exhibits a standing wave feature. However, at the centre of the wide bridge, the 5-min oscillation power is found to be stronger than at its sides. Moreover, the time-distance plot across the wide bridge exhibits a herringbone pattern that indicates a counter-stream of two running waves originated at the bridge sides. 
 Thus, the 5-min oscillations on the wide bridge also resemble the properties of running penumbral waves. The 5-min oscillations are suppressed in the umbra, while the 3-min oscillations occupy all three cores of the sunspot's umbra, separated by the LBs. The 3-min oscillations were found to be in phase at both sides of the LBs. It may indicate that either LBs do not affect umbral oscillations, or umbral oscillations at different umbral cores share the same source. Also, it indicates that LBs are rather shallow objects situated in the upper part of the umbra. We found that umbral flashes follow the life cycles of umbral oscillations with much larger amplitudes. They cannot propagate across LBs. Umbral flashes dominate the 3-min oscillation power within each core, however, they do not disrupt the phase of umbral oscillation.
\end{abstract}
\keywords{Sun: atmosphere --- Sun: chromosphere --- Sun: oscillations --- sunspots --- magnetohydrodynamics (MHD) --- waves}

\section{Introduction}
\label{sec:intro}
Waves and oscillations in sunspots are one of the most extensively studied magnetohydrodynamic (MHD) wave phenomena in solar physics \citep[see a comprehensive review by][]{bogdan2006}. The associated MHD seismology is a potential tool to probe a sunspot's thermal and magnetic structure \citep[e.g.,][]{zhugzhda1983, zhugzhda2008,shibasaki2001,yuan2014cf}, and photospheric-coronal magnetic connectivity \citep{sych2009,yuan2011lp}. The oscillation power distribution of different periods in a sunspot is non-uniform in both horizontal and vertical directions. The 3-min oscillations dominate a sunspot's umbra \textbf{in the chromosphere}, while the 5-min oscillations are most prominent in the penumbra \citep[see, e.g.,][]{bogdan2006,yuan2014cf}.

Umbral oscillations are usually interpreted as standing slow mode magnetoacoustic waves \citep[e.g.,][]{christopoulou2000,christopoulou2001,botha2011}. The emission intensity and Doppler velocity inside sunspot umbrae oscillate collectively with a period of about 3-min, reaching the maximum amplitudes at the chromospheric heights  \citep[e.g.,][]{reznikova2012a,yuan2014cf}. The effect of height inversion occurs in the umbral oscillation power: a hump in the oscillation power is usually found at chromospheric heights, however, a power depletion is usually detected at photospheric heights underneath \citep{kobanov2011}. \citet{aballe1993} found a clear correlation between 3-min oscillation power and umbral brightness in the dark core. In the corona, the 3-min oscillations become propagating slow magnetoacoustic waves and follow the magnetic fan structures extending out from the sunspot \citep{demoortel2002a,demoortel2002b,demoortel2009,botha2011,yuan2012sm,kiddie2012}.

Umbral flashes \citep[UFs,][]{beckers1969} are seen as strong brightenings occurring at seemingly random locations in sunspot umbrae. Their repetition rate is about 2--3 min, indicating their possible connection with umbral oscillations  \citep{roupevandervoort2003,delacruzrodriguez2013}. Chromospheric UFs are proposed to be magnetoacoustic shock fronts produced by the photospheric sound waves \citep{havnes1970}, that is consistent with the sawtooth waveform of the UF trains \citep[see e.g.,][]{tian2014}. \citet{bard2010} performed 1D radiative hydrodynamic simulations, and demonstrated that UFs are enhanced emissions of local plasma during the passage of photospheric acoustic waves. High-resolution spectrometric studies appear to support this theory  \citep[e.g.,][]{tziotziou2002,tziotziou2006,tziotziou2007,roupevandervoort2003,delacruzrodriguez2013}. \citet{socas2000a,socas2000b} detected anomalous Stokes spectra in conjunction with UF occurrences and claimed that UFs were associated with hot up-flowing material and a rest cool component. Moreover, \citet{socas2009} found rich fine structures of umbral flashes and measured unusually high lateral propagating speed that exceeded local fast speed, excluding the naive 
explanation of UF in terms of  propagating fast magnetoacoustic waves. Thus, there remain a number of open questions associated with the specific details of the UF physics. 

Longer-period spectral components, commonly known as 5-min oscillations, are usually suppressed inside umbrae \citep{zirin1972}. Significant oscillation power forms a ring-structure at the umbra-penumbra boundary, with the radius of the rings increasing with the increase in the oscillation period \citep{nagashima2007, sych2008, reznikova2012a, yuan2014cf}. The physical mechanism responsible for such a behaviour is still unclear. Solar $p$-mode acoustic wave is a candidate energy source \citep{abdelatif1986, jain2009, parchevsky2009}. In particular, the interaction of $p$-modes with the strong magnetic field in a sunspot could excite magnetoacoustic modes \citep{cally1997, cally2003, schunker2006, khomenko2013}. \citet{penn1993} suggested that $p$-mode absorption by sunspots occurs linearly across a sunspot umbra or within a ring surface where local magnetic field allows for the optimised absorption rate. Moreover, such ring-surface absorption is favoured in theoretical studies \citep{cally2003, schunker2006}, with the $p$-mode absorption found to be optimal at an attack angle of about $30$ degrees. 

Penumbral magnetic field deviates gradually from the solar normal with the distance from the umbra-penumbra border and becomes almost horizontal at the supra-penumbra \citep{borrero2011, solanki2003}. Thus, the relatively simple magnetic geometry of the umbral and penumbral magnetic field allow for the development of relatively simple models describing sunspot oscillations and their application in observations. However, there are fine details such as umbral dots and light bridges (LBs) in umbrae. LBs are usually seen as bright extended filaments across the umbrae. The magnetic structure of LBs is believed to be rather different from the almost vertical and uniform field in the umbrae. It is likely to form a magnetic canopy with a large horizontal component \citep{rueedi1995, leka1997, jurcak2006}. Sunspot oscillations have not been investigated against such magnetic topology. However, bright umbral structures, such as dots and LBs, were demonstrated numerically to deteriorate sunspot oscillations and hence reveal further physics \citep{locans1988}. Recently, \citet{sobotka2013} performed analysis of oscillations in a pore with an LB, but without a penumbra, and found that the oscillations in the LB were very different from the usual 3-min oscillations in the umbral part of the pore.

In this study, we present a high-cadence and fine-resolution observation of oscillations in a well-developed sunspot with two LBs of different width, at the chromospheric level. We focus on the spatial distribution of 3-min, including UFs, and 5-min oscillations in the sunspot's umbra and LBs, investigating how LBs affect umbral oscillations. We present data preparation in \secref{sec:obs}, analysis and results in \secref{sec:anal}, and conclusions in \secref{sec:con}.

\section{Observations}
\label{sec:obs}

\begin{figure}[ht]
\centering
\includegraphics[width=0.4\textwidth]{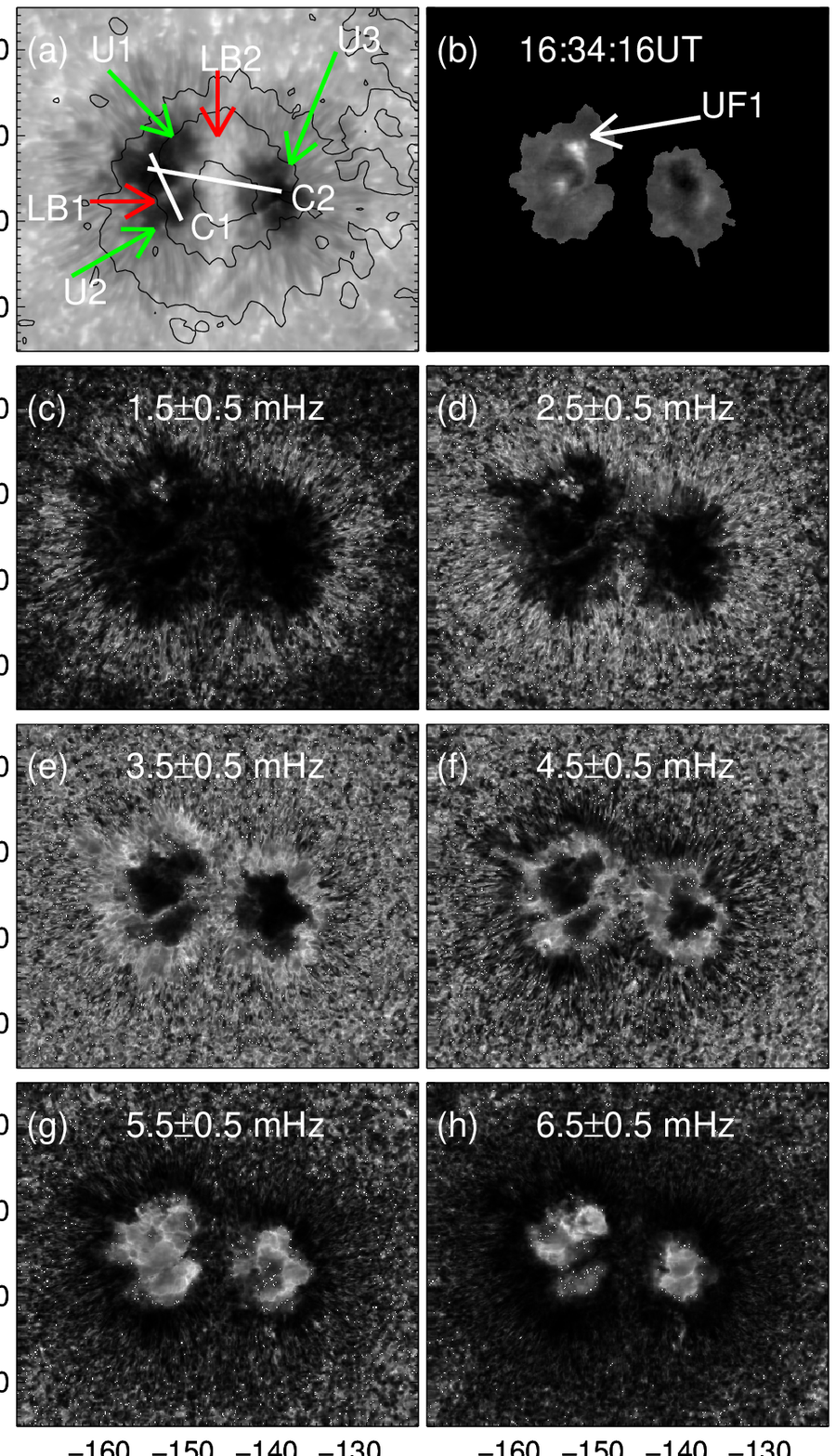}
\caption{(a): The SOT \ion{Ca}{2} H emission intensity image of \arobs, in a logarithmic scale. A faint and strong light bridges (LB1 and LB2) divide the sunspot umbra into three parts: U1, U2 and U3. HMI magnetogram (contour) illustrates that all three parts of the umbra are of the same magnetic polarity. Two cuts across the light bridges, used in the following analysis, are annotated as C1 and C2. (b) A \ion{Ca}{2} H difference image exhibiting the spatial extent of UF1, only the umbral region is shown. (c)-(h) are the normalised narrowband Fourier power maps averaged over 1 mHz bands around the central frequencies. The central frequencies of the spectral bands are labeled in each panel. \label{fig:fov}}%
\end{figure}

\begin{figure}[ht]
\centering
\includegraphics[width=0.40\textwidth]{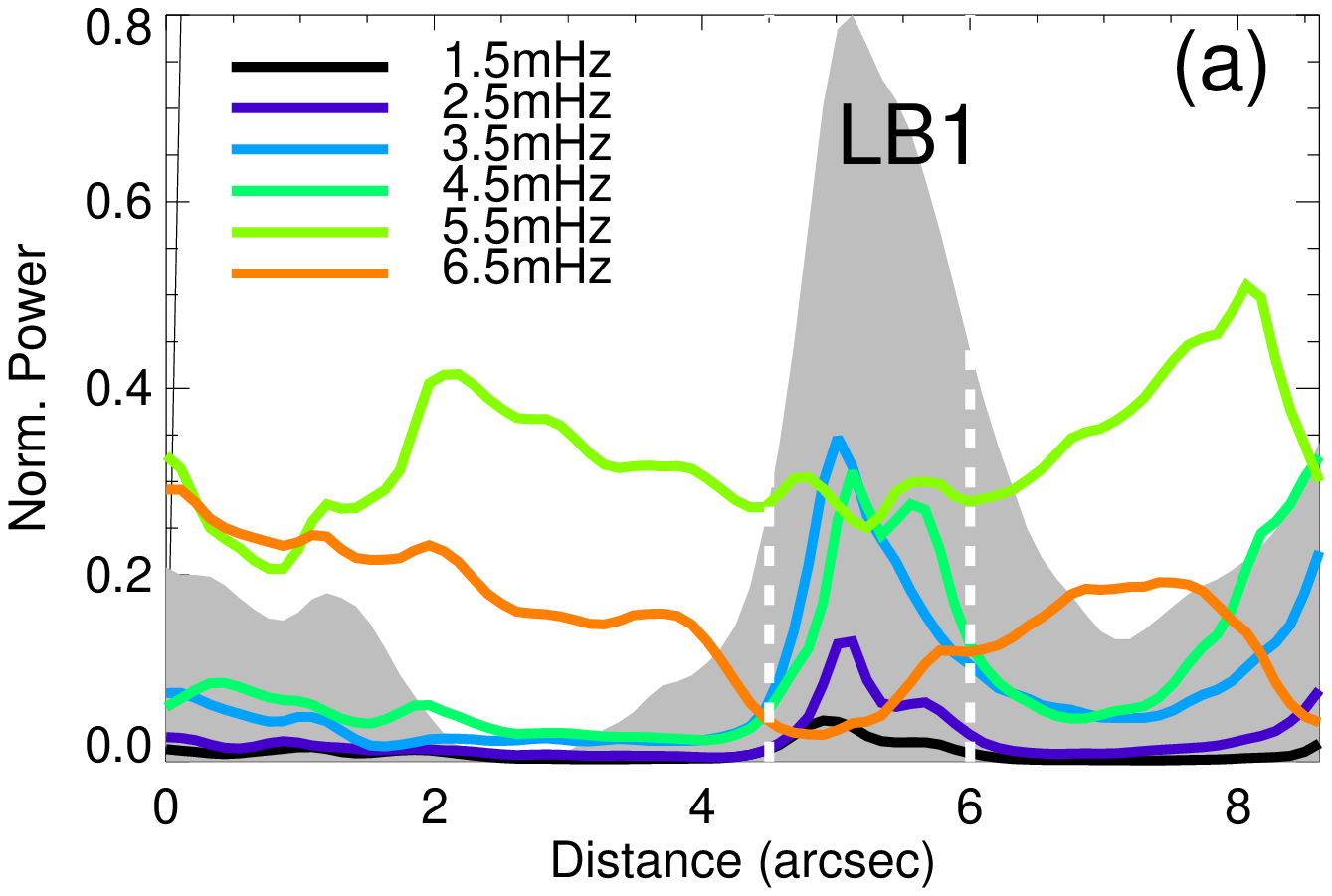}
\includegraphics[width=0.40\textwidth]{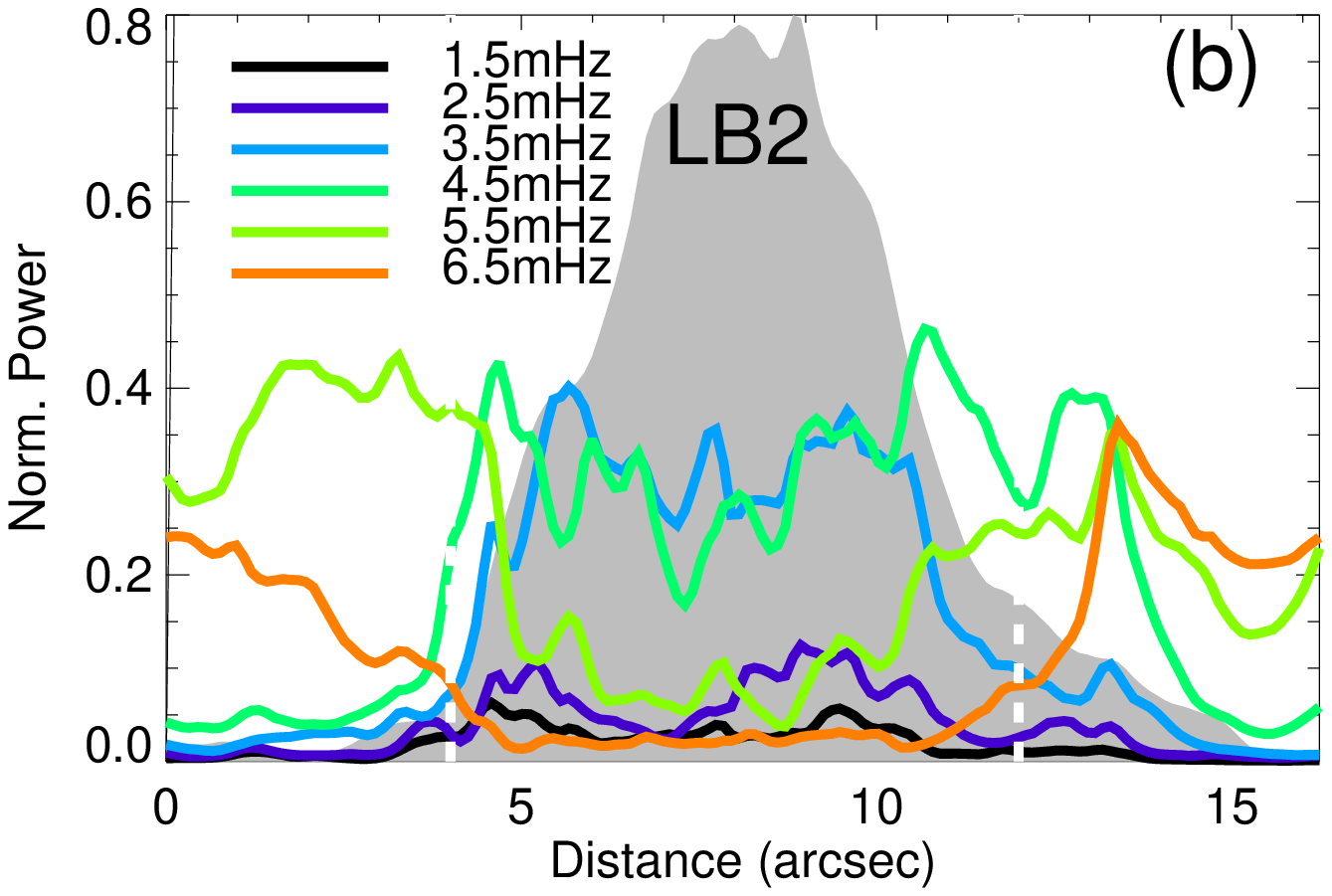}
\caption{The profiles of the oscillation power profiles along cuts C1 (a) and C2 (b) for different frequency bands. The positions of LB1 and LB2 are shown by the scaled intensity variations (grey shade) along C1 and C2 \label{fig:prof}}%
\end{figure}

The Solar Optical Telescope \citep[SOT,][]{tzuneta2008} onboard the Hinode satellite \citep{kosugi2007} observed the chromosphere of a divided sunspot \arobs on \dateobs. A faint and a strong light bridge (labeled as LB1 and LB2, respectively) divided the umbra into three cores of the same magnetic polarity (U1, U2 and U3, see \figref{fig:fov}a). LB2 was a strong photospheric (granular) bridge \citep{muller1979,sobotka1994} and was observed in SOT's both \ion{Ca}{2} H (3968.5 \AA{}) bandpass and G-band\footnote{G-band observations over \arobs do not have good cadence for this study, thus they are not used.}. It was formed before \arobs became first observable on the solar disk on 28 August, faded off accompanying the shrinking of umbral core U3, and disappeared on 02 September, according to the continuous observation with the Solar Dynamic Observatory \citep[SDO,][]{pesnell2012}. It is a typical splitting process when a main spot is crossed by a wide granular bridge \citep{vazquez1973}. LB1 was an umbral streamer \citep{muller1979}, also visible in SOT's both bandpasses, however, its evolution was hardly resolved with SDO.

The SOT's Broadband Filter Imager (BFI) obtained a series of $512\times512$ pixels filtergrams in the \ion{Ca}{2} H bandpass from 16:00 to 18:00 UT on \dateobs. The cadence was about 15 s and the pixel size was approximately $0.1$ arcsec. The SolarSoft routine \textit{fg\_prep.pro} calibrated the fits images by subtracting the dark current, applying a flat field, correcting the bad pixels, and normalising them with their exposure times. The SOT images were then reframed onto the Helioseismic and Magnetic Imager \citep[HMI onboard SDO,][]{schou2012} image coordinates by matching the sunspot centers. We tracked a $48\arcsec\times40\arcsec$ region against the solar differential rotation and obtained a sequence of images of $441\times368$ pixels in size. The images were co-aligned by compensating the offsets obtained by the cross-correlation technique and were interpolated into a uniform time grid with a 15 s cadence. We detrended and normalised the images with 20 points running mean values and obtained a baseline-ratio difference image set \citep[see the technique details in, e.g.,][]{aschwanden2011, yuan2012sm}.

\section{Analysis and results}
\label{sec:anal}
We first examined the spatial distribution of the oscillation power over the sunspot. An FFT transform was applied pixel-by-pixel to the baseline-ratio difference array. Time series of each pixel was apodised with a Tukey window with $\alpha=0.2$ \citep{harris1978} before FFT to mitigate the effect of a finite observation interval. The obtained power maps were normalised with their maximum power. \figref{fig:fov}c-h display the normalised narrowband power maps averaged over 1~mHz bands and centred at the frequencies of 1.5, 2.5, $\cdots$, and 6.5~mHz. \figref{fig:fov} shows clearly that high-frequency oscillations fill the umbrae, while the low-frequency counterparts take up the penumbra. This result is consistent with previous studies \citep[e.g., ][]{nagashima2007, reznikova2012b,jess2013,yuan2014cf}.

\subsection{Oscillations on light bridges}

LBs are seen to be filled in with the oscillation power at frequencies of 3.5 mHz and 4.5 mHz (the 5-min band, \figref{fig:fov}), but devoid of 5.5 mHz and 6.5 mHz oscillation power (the 3-min band). To illustrate this effect, we took the power profiles along cuts C1 and C2 that cross LB1 and LB2, respectively (see \figref{fig:prof}).  It is clear that oscillation powers of 3.5 mHz and 4.5 mHz exhibit humps in the LBs, while those of 5.5 mHz and 6.5 mHz are depleted in the LBs. LBs in pores were found to exhibit a similar feature \citep{sobotka2013}. 

\begin{figure*}[ht]
\centering
\includegraphics[width=0.40\textwidth]{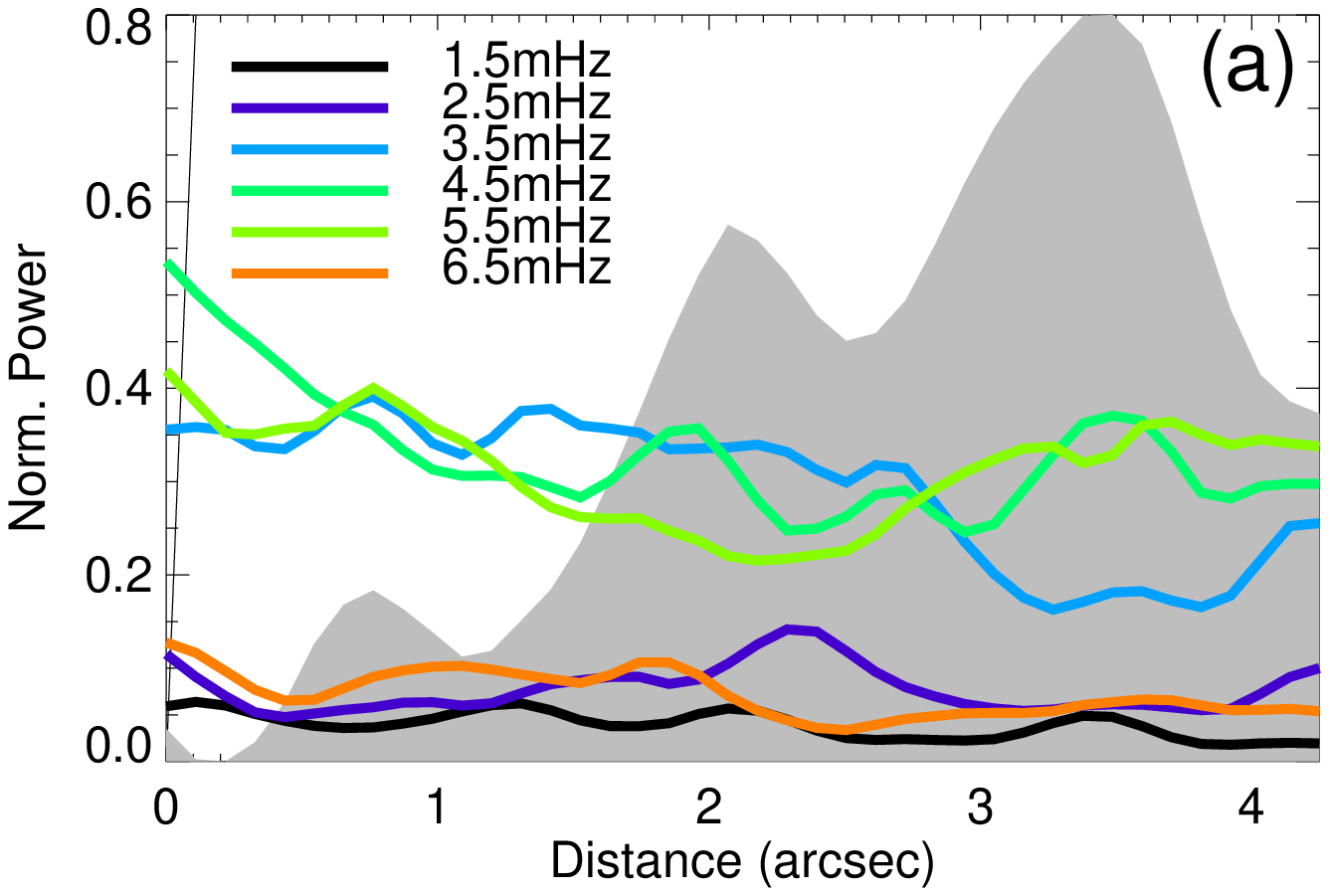}
\includegraphics[width=0.20\textwidth]{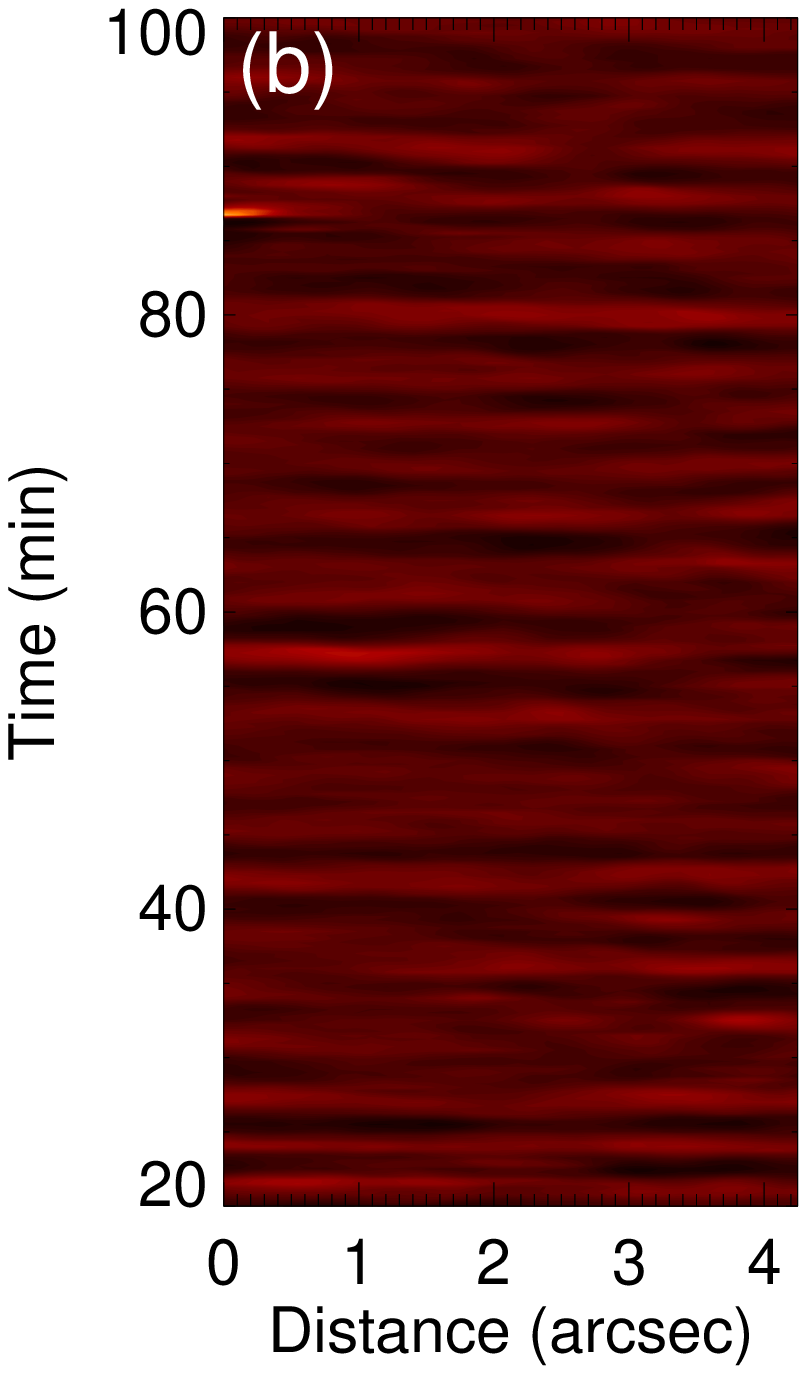}
\caption{(a) The narrowband power profiles along LB1 in different frequency bands. \textbf{The grey shade denotes the emission intensity variation along the axis of the bridge.} (b) The time-distance plot along LB1 in the baseline-difference array. \label{fig:lb1}}%
\end{figure*}

Sources of 5-min oscillations on the faint bridge supposedly are either laterally at the penumbral ends of the bridge or come from underneath. In the former case, we expect an edge-to-centre variation of 5-min oscillation power and phase along the bridge. We took the power profiles of the oscillations along LB1, shown in \figref{fig:lb1}a. No clue proves that the bridge ends have stronger 5-min oscillations than its centre. Moreover, the time-distance plot made along LB1 (\figref{fig:lb1}b) reveals no propagating features. Therefore, the 5-min oscillations are more likely to originate from underneath, rather than coming on the bridge from its ends anchored in the penumbra, and exhibit features of standing slow magnetoacoustic wave

\begin{figure}[ht]
\centering
\includegraphics[width=0.30\textwidth]{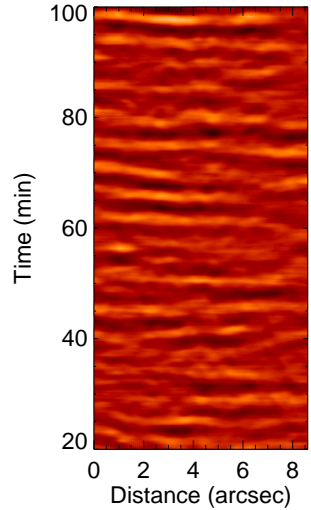}
\caption{The time-distance plot along LB2 in the baseline-difference array. \label{fig:lb2}}%
\end{figure}

LB2 is stronger and wider than LB1, thus its fine structure may reveal more physics. The time-distance plot along LB2 (\figref{fig:lb2}) exhibits similar standing wave features as observed along LB1. The LB2 part of the time-distance plot along C2 (normal to LB2, \figref{fig:cuts}b) exhibits a weak herringbone pattern, indicating a counter-stream of running waves from two umbral cores U1 and U3. The apparent propagation speed is about $10-20\,\textrm{km}/\textrm{s}$.
 Moreover, the 3.5 and 4.5 mHz oscillation power is slightly lower at the LB2's centre than those at its sides (\figref{fig:prof}b). We observe waves propagating inwardly from the bridge long sides to its central axis. In the direction along the central axis of the bridge, we see a standing wave.

\subsection{Oscillations around light bridges}
\label{sec:umbra}

\begin{figure*}[ht]
\centering
\includegraphics[width=0.3\textwidth]{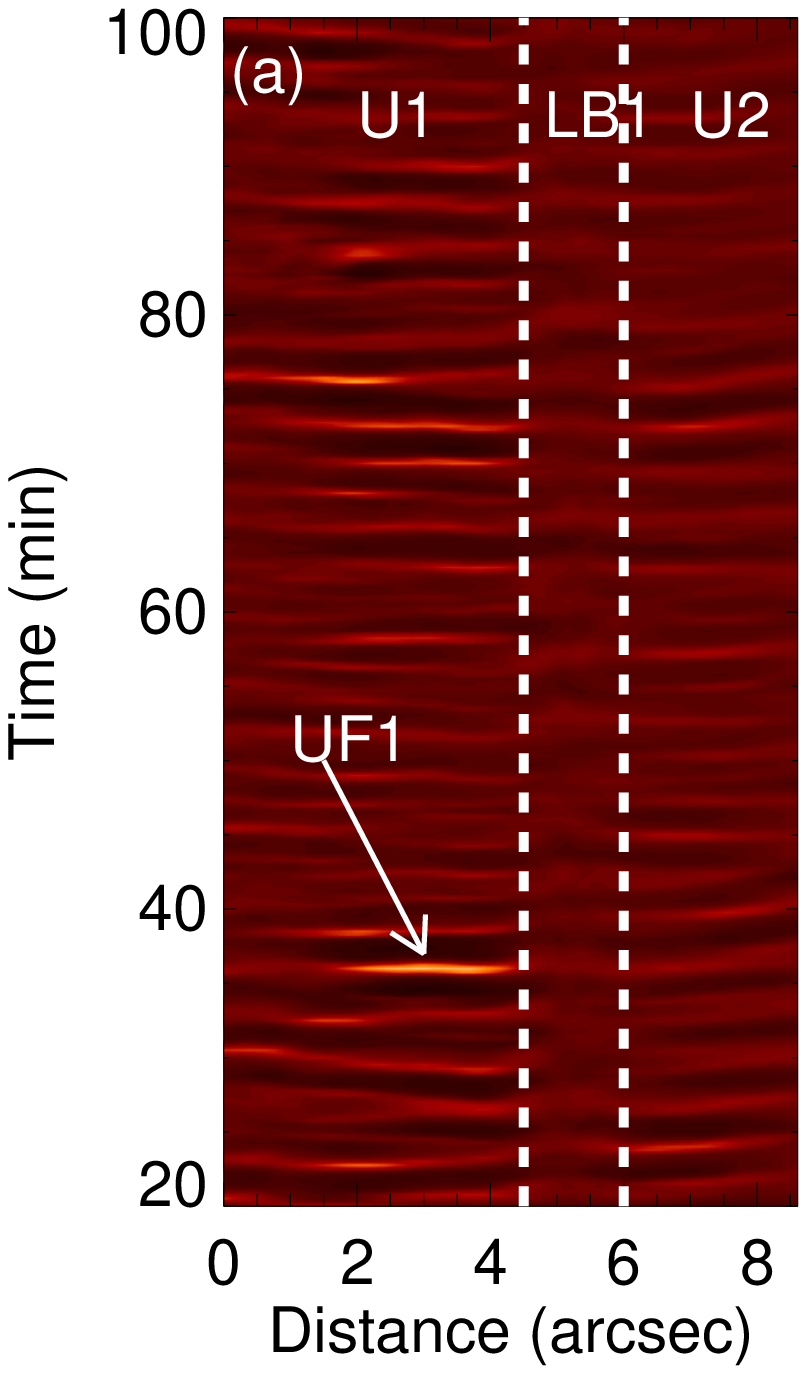}
\includegraphics[width=0.3\textwidth]{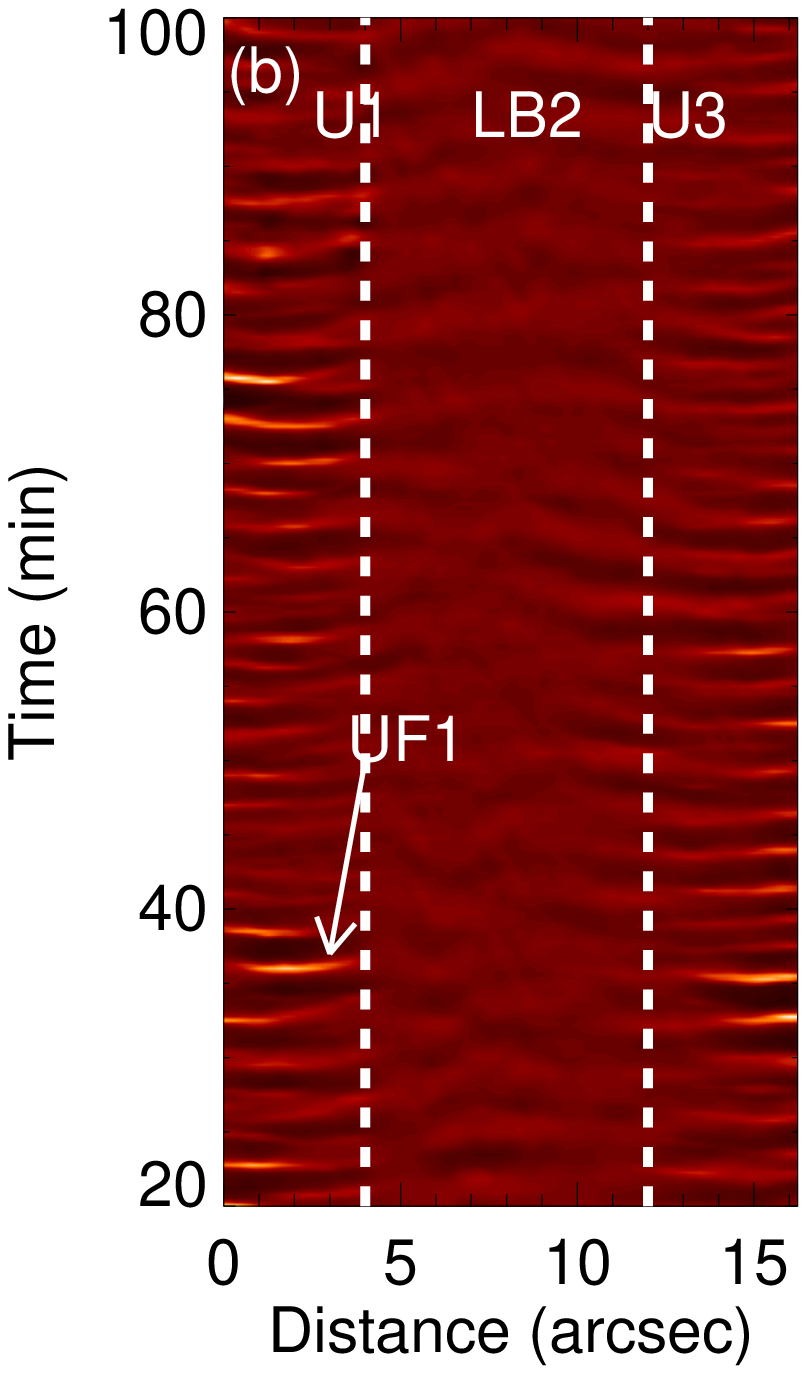}
\caption{The time-distance plot of cuts C1 (a) and C2 (b), the approximate positions of LB1 and LB2 are enclosed by dashed lines. UF1 is labeled in each panel. \label{fig:cuts}}%
\end{figure*}

\begin{figure}[ht]
\centering
\includegraphics[width=0.40\textwidth]{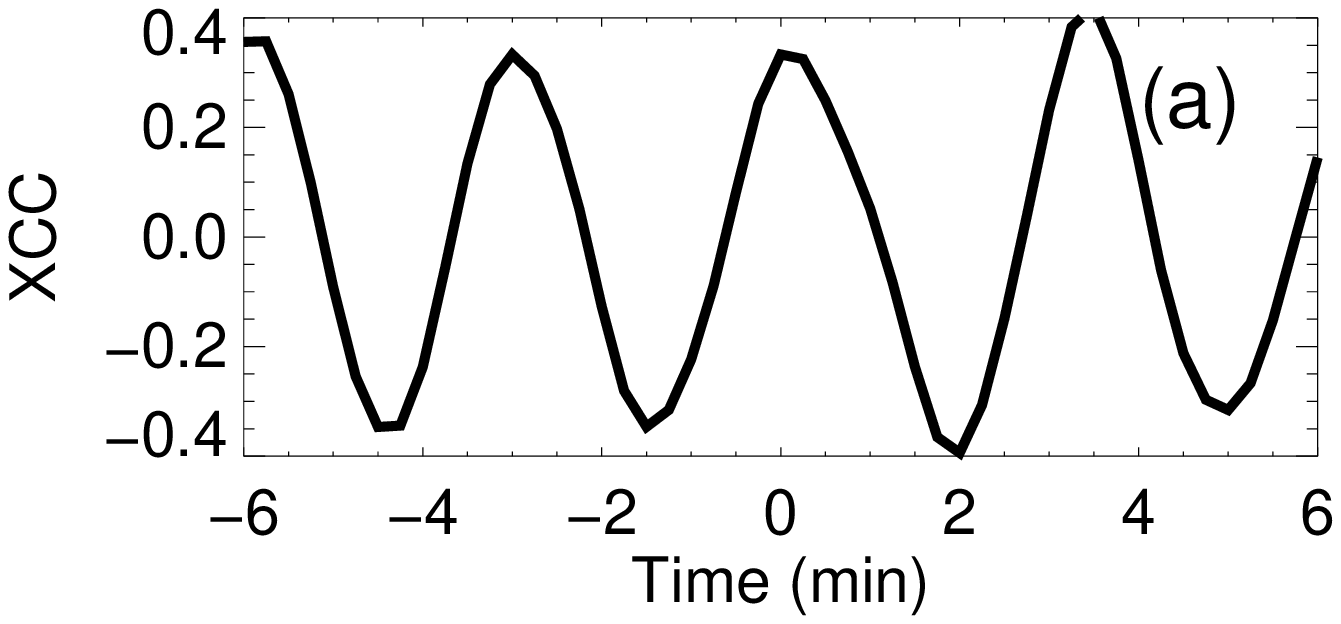}
\includegraphics[width=0.40\textwidth]{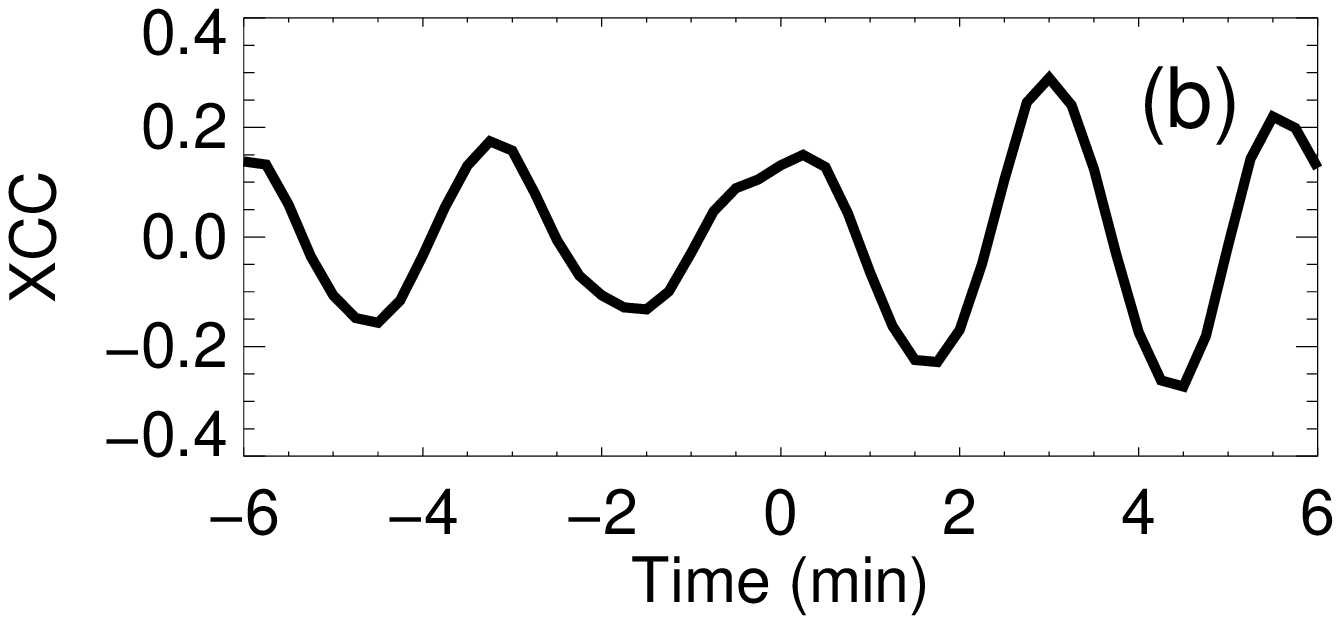}
\caption{The cross-correlation coefficient (XCC) of the time series averaged at two sides of LB1 (a) and LB2 (b), respectively, as a function of lag time. \label{fig:lag}}%
\end{figure}

The 3-min oscillations were \textbf{delimited by} umbral boundaries and LBs (\figref{fig:fov}). \figref{fig:cuts} displays a segment of the time-distance maps along C1 and C2 in the baseline-ratio difference array. The horizontal ridges seen in the time-distance plots are well-consistent with the interpretation in terms of standing slow magnetoacoustic waves \citep[see, e.g.,][]{christopoulou2000,christopoulou2001, botha2011}. The wave fronts on both sides of a LB appear to have almost a one-to-one correspondence. To quantify this effect, we selected the U1 and U2 parts (located at $0-4.0\arcsec$ and $6.0\arcsec-8.6\arcsec$, respectively) of the time-distance array along cut C1 (see the dashed lines in \figref{fig:cuts}a), and obtained two time series by averaging the selected partial time-distance arrays over the space domain. We calculated the cross-correlation coefficient (XCC) of the two time series for different lag times, see \figref{fig:lag}a. Thresholding was applied at a value of 0.2 to mitigate the effect of the spikes introduced by UFs. The first maximum of the XCC was detected at a zero lag time. It means that the umbral oscillations at both sides of LB1 are apparently in phase. We applied the same analysis to the U1 ($0-4\arcsec$) and U3 ($12\arcsec-16.4\arcsec$) parts of the time-distance map along C2 and obtained that the time lag corresponding to the maximum correlations is about 15 sec. As this value is about the cadence time of the measurements, see \figref{fig:lag}b, and much shorter than the oscillation period, hence it could also be considered as a zero lag time. Therefore, our analysis reveals that the umbral oscillations in U1, U2 and U3 are apparently in phase, although the spatial locations of these oscillations are separated by LBs. 

\subsection{Umbral flashes}

\begin{figure*}[ht]
\centering
\includegraphics[width=0.8\textwidth]{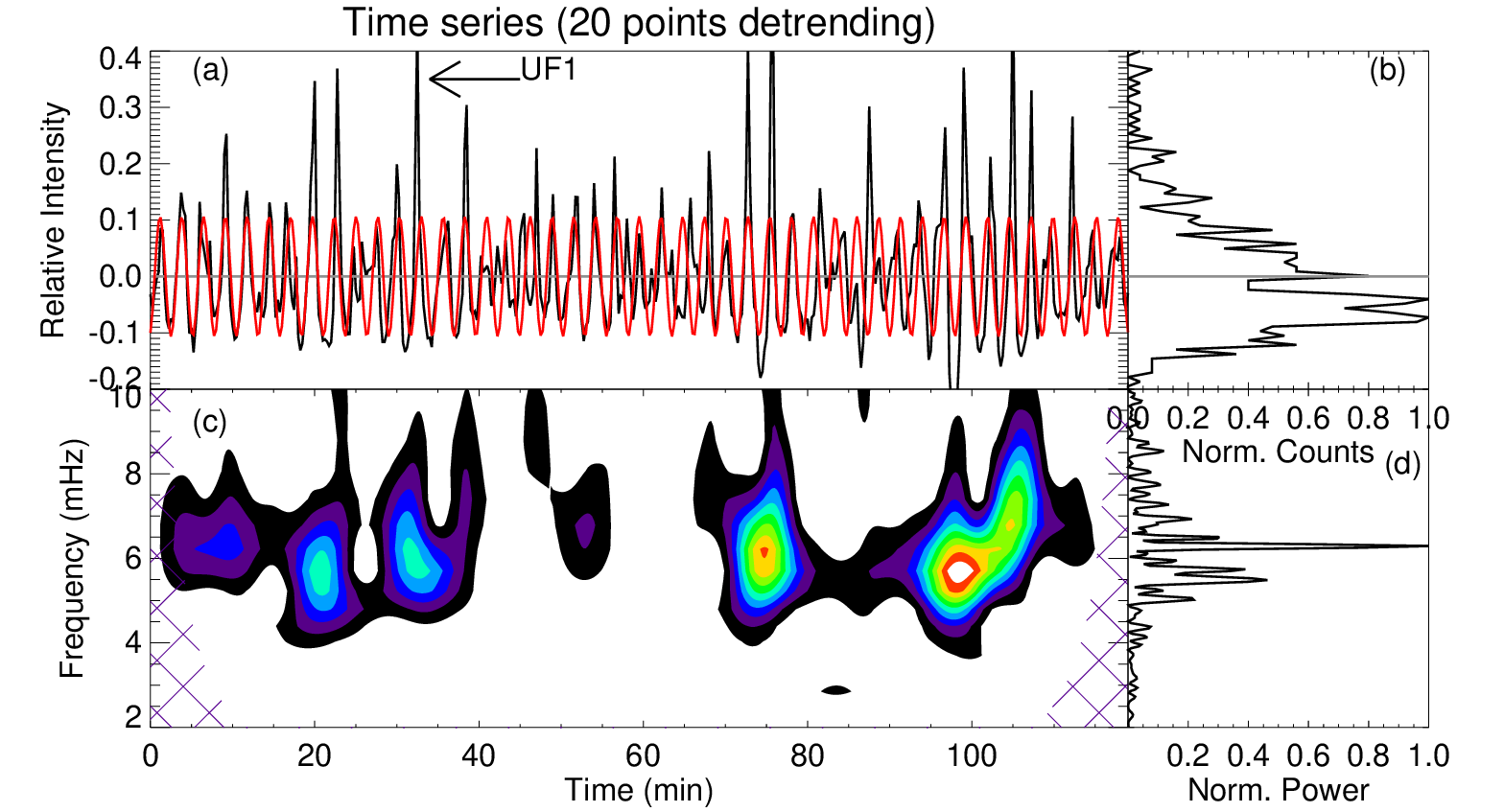}
\caption{(a): The baseline-ratio intensity variation of a pixel taken at U1. The red curve is a single harmonic fit to the time series.  Other panels are (b) the histogram of time series, (c) the Morlet wavelet spectrum, and (d) the periodogram. \label{fig:wavelet}}%
\end{figure*}
Apart from the quasi-monochromatic oscillations discussed above, the observed sunspot hosts interesting examples of UF phenomenon. UFs are seen as local emission intensity enhancements in chromospheric umbrae (see \figref{fig:fov}b) of the analysed sunspot. This phenomenon is illustrated by UF1 as an example: its spatial extent (\figref{fig:fov}b), its spatiotemporal  morphology (\figref{fig:cuts}), and its time profile (\figref{fig:wavelet}a).

We see that UFs occur as trains of several sharp increases in the brightness. Inside the trains, the UFs repeat with a 3-min periodicity that is consistent with the early findings \citep[see, e.g.,][]{roupevandervoort2003, delacruzrodriguez2013}. \figref{fig:cuts} reveals that the UF trains occur at random locations without a well-established occurrence rate. Moreover, individual UFs are seen to ride wave fronts of umbral oscillations. But, in contrast to the apparent coherence of the 3-min oscillations on either sides of LBs discussed in Sec.~\ref{sec:umbra}, UFs lack the one-to-one correspondence at either sides of LBs (see \secref{sec:umbra}). 
Thus, UFs are confined within the umbral cores and cannot propagate across LBs. 

\figref{fig:wavelet} plots the time series of a pixel in the U1 part of C2 and its power spectrum. The amplitude of umbral oscillations is normally less than 10\% of the background intensity, but surges up to 60\% when an UF occurs. The series of spikes introduced by UFs constitute the major oscillation power in the wavelet spectrum (\figref{fig:wavelet}c) and contribute significantly to the peak at about 6.2 mHz in the periodogram (\figref{fig:wavelet}d). Between UFs, umbral oscillations produce minor power in comparison with UF oscillations in spite of their persistence over the whole time series.

The time series of the intensity variation were over-plotted with a sinusoidal fit (red curve, \figref{fig:wavelet}a). This figure confirms the finding that UFs follow the cycles of umbral oscillations (also see \figref{fig:cuts}) and that UFs do not disrupt the phase of 3-min oscillations. Thus, we see that UFs' intensity variation (\figref{fig:wavelet}a) exhibits no significant difference with 3-min oscillations in the umbra, apart from having a much larger amplitude, at about 50\% of the background intensity. 

\section{Conclusions}
\label{sec:con}

We have analysed the intensity variations in a sunspot \arobs with two LBs, observed with Hinode/SOT in the \ion{Ca}{2} H bandpass. In full agreement with the commonly accepted knowledge, the 3-min oscillations were found to occupy the umbral part of the sunspot, while 5-min oscillations fill in the penumbra. Our narrowband power map analysis shows that significant 5-min oscillations are also present in LBs, while 3-min oscillations are suppressed there. Our finding generalises the earlier results obtained recently by \citet{sobotka2013} for a pore for the case of a well-developed sunspot with a penumbra. The 5-min oscillation power along the faint bridges (LB1) does not exhibit recognisable feature, however, the time-distance plot along LB1 illustrated that the 5-min oscillations do not experience a noticeable change in the phase along the bridge. Thus, 5-min oscillations are found to exhibit a standing wave behaviour along the thin bridge (\figref{fig:lb1}), the same process was found along the central axis of LB2 (\figref{fig:lb2}).

We should point out that there could be a concern that the contamination of stray light would generate a spurious signal and affect the subsequent analysis, since LB1 is measured to be less than $2\arcsec$ wide and about $4\arcsec$ long. In the discussed case this effect is negligible, since LB1 was surrounded by umbrae that were much darker. Moreover, if the stray light did contribute significantly to the emission of LB1, we would expect to observe significant 3-min oscillations associated with the umbral oscillations, in the LB. However, our analysis did not show the presence of 3-min oscillations in the LBs. 

The 5-min oscillations on the strong bridge (LB2) show a rather complex behaviour. The time-distance plot across LB2 exhibits a weak herringbone pattern that denotes  two waves running towards each other from the opposite sides of the bridge. The projected propagation speed is about $10-20\,\textrm{km}/\textrm{s}$. As this value is about the typical values found for running penumbral waves, and as the waves are seen to propagate in the direction perpendicular to the LB sides, it may be considered as an LB counterpart of the running penumbral wave phenomenon.This finding may indicate that the wave carries energy from the umbral cores inwards the LB. But, one should be cautious with this interpretation, as the apparent wave motion may not be associated with any energy transfer. Indeed, what is measured is the phase speed that does not represent any energy transfer, and the apparent wave motion can be caused by, e.g., the projection of the waves propagating along the magnetic field lines. Our finding demonstrates the need for a more detailed study of this effect.

The suppression of 3-min oscillations in the LB can be associated with the departure of the magnetic field lines from the vertical over the bridge. Indeed, if in the chromosphere above a LB the field has a canopy geometry (see, e.g., the sketch shown in Fig.~7 of \citet{jurcak2006}). Thus the LB effect on MHD waves should be similar to this of the umbra-penumbra boundary, and hence the resonant properties of the LB and of the umbra-penumbra boundary should be similar. This results in the observed suppression of 3-min oscillations and appearance of 5-min oscillations at both LB and umbra-penumbra boundaries. However, there is a important issue connected with the excitation of the 5-min oscillations in LBs. The well-developed theory of the $p$-mode absorption by sunspots \citep{cally2003,schunker2006,jain2009,khomenko2013} should answer the question how the acoustic oscillations come to the LB. To illustrate this problem, consider a LB as the top of a vertical magnetic-free (or, almost magnetic-free) slab surrounded by the strong umbral magnetic field. The top boundary of the slab is formed by the magnetic canopy. The $p$-modes surrounding the sunspot cannot penetrate the LB through the long sides formed by the umbral magnetic field. Thus, there are two possibilities, either the p-modes enter the LB through its ends, or from underneath. In the either case, the angular spectrum of the $p$-modes that can reach the LB is very restricted. This should possibly be somehow seen in the observations, either as a significant decrease in the 5-min oscillation power in the LB in comparison with the umbra-penumbra boundary, or as formation of waves propagating along the bridge and carrying the 5-min energy to its centre. Neither effect was detected by our analysis in the thin bridge: the spatial structure of 5-min oscillations at the umbra-penumbra boundary and at the LBs is similar. But, there is a clear evidence of the waves propagating from the sides of the wide bridge towards its centre. Thus, modelling of the $p$-mode interaction with a sunspot with an LB is an interesting topic for theoretical consideration.

In addition, our analysis showed that 3-min umbral oscillations at both sides of both the wide and thin LBs are in-phase. Perhaps, this effect is associated with the excitation of the 3-min umbral oscillations on the either sides of the LB by the same source that apparently \lq\lq does not feel\rq\rq the LB, or with some cross-talk of the 3-min oscillations occurring at the neighbouring (while separated by the LB) spatial locations. It implies that light bridges are shallow objects situated in the upper part of the umbra. The umbral cores are probably connected below the LB surface. In this scenario, the standing waves observed in bridges are not connected with $p$-modes, as they are not likely to reach the bridges because of the surrounding umbra. It suggests that 5-min standing oscillations in light bridges are possibly standing acoustic oscillations trapped in the vertical non-uniformity of the plasma density and temperature along the magnetic field \citep{zhugzhda2008, botha2011}. Also, the possibility of just a coincidence cannot be ruled out, and more observational examples need to be studied to assess the statistical significance of this finding.

The parts of the umbra separated by LBs showed transient UFs. The UFs were found to follow the cycles of umbral 3-min oscillations and do not disrupt their phase. Thus, UFs are seen to be high amplitude parts of the amplitude-modulated 3-min oscillations \citep[see][for the discussion of the modulation of 3-min umbral oscillations]{sych2012}. In UFs the amplitude of 3-min oscillations reaches 60\% of background intensity, and dominates in the 3-min spectral peak. The finite-amplitude (weakly nonlinear) effects naturally lead to the wave steepening and formation of acoustic shocks. UFs were found to be constrained within an umbral core. In contrast with 3-min oscillations discussed above, UFs on the either sides of LBs did not show significant correlation.

\begin{acknowledgements}
This work is supported by the Marie Curie PIRSES-GA-2011-295272 \textit{RadioSun} project, the European Research Council under the \textit{SeismoSun} Research Project No.~321141 (DY, VMN), the Open Research Program KLSA201312 of the Key Laboratory of Solar Activity of the National Astronomical Observatories of China (DY), the Russian Foundation of Basic Research under grant 13-02-00044; the BK21 plus program through the National Research Foundation funded by the Ministry of Education of Korea (VMN), the National Natural Science Foundation of China (40904047, 41174154, and 41274176), the Ministry of Education of China (20110131110058 and NCET-11-0305), the Provincial Natural Science Foundation of Shandong via Grant JQ201212 (BL, DY), the China 973 program 2012CB825601, NSFC Grants 41274178 (ZHH), 11373040 (JTS), 11273030, 11221063, and MOST Grant 2011CB811401 (BLT, YHY).
\end{acknowledgements}

{\it Facility:} \facility{Hinode (SOT)}

\bibliographystyle{aa}

\end{document}